\documentclass[preprint,12pt]{elsarticle}

\usepackage{color}
\usepackage{ulem}

\usepackage{graphics}

\usepackage{amssymb}
\usepackage{todonotes}
\usepackage{hyperref}
\usepackage{xspace}

\newcommand{\be}{\begin{equation}}
\newcommand{\ee}{\end{equation}}
\newcommand{\bea}{\begin{eqnarray}}
\newcommand{\eea}{\end{eqnarray}}

\newcommand{\pth}{\texttt{PArthENoPE}\xspace}
\newcommand{\pthdue}{\texttt{PArthENoPE 2.0}\xspace}
\newcommand{\pthnew}{\texttt{PArthENoPE 3.0}\xspace}

\newcommand{\nn}{\nonumber}

\newcounter{bla}

\begin{document}

\begin{frontmatter}

\title{\pth Revolutions}


\author[a,b]{S. Gariazzo}
\author[c]{P.\ F.\ de Salas}
\author[d]{O. Pisanti\corref{pisanti}}
\author[d]{R. Consiglio}

\cortext[pisanti] {Corresponding author.\\\textit{E-mail address:} pisanti@na.infn.it}

\address[a]{INFN, Sezione di Torino, Via P. Giuria 1, I--10125 Torino, Italy}
\address[b]{Instituto de F{\'\i}sica Corpuscular (CSIC-Universitat de Val{\`e}ncia),
Parc Cient{\'\i}fic UV, C/ Ca\-te\-dr{\'a}tico Jos{\'e} Beltr{\'a}n, 2, E-46980 Paterna (Valencia), Spain}
\address[c]{The Oskar Klein Centre for Cosmoparticle Physics,
Department of Physics, Stockholm University, SE-106 91 Stockholm, Sweden}
\address[d]{Dipartimento di Fisica E. Pancini, Universit\`a di Napoli Federico II, and INFN, Sezione di Napoli, Via Cintia, I-80126 Napoli, Italy.}

\begin{abstract}
This paper presents the main features of a new and updated version of the program \pth, which the community has been using for many years for computing the abundances of light elements produced during Big Bang Nucleosynthesis.
This is the third release of the \pth code, after the 2008 and the 2018 ones, and
will be distributed from the code's website, \url{http://parthenope.na.infn.it}.
Apart from minor changes, the main improvements in this new version include a revisited implementation of the nuclear rates for the most important reactions of deuterium destruction, $^2$H(p,$\gamma)^3$He, $^2$H(d, n)$^3$He and $^2$H(d, p)$^3$H, and a re-designed GUI, which extends the functionality of the previous one. The new GUI, in particular, supersedes the previous tools for running over grids of parameters with a better management of parallel runs, and it offers a brand-new set of functions for plotting the results.
\end{abstract}


\end{frontmatter}



%

\section{Introduction}
\label{s:intr}

Nowadays, the success of the standard cosmological model ($\Lambda$CDM), which includes a radiation dominated epoch during the first stages after the Big Bang, is beyond doubt. One of the crucial processes during this epoch is the Big Bang Nucleosynthesis (BBN), which 
explains the formation of the first light nuclei (mainly $^{2}$H, $^{3}$He, $^{4}$He and $^{7}$Li).
The abundances of such elements established during BBN can be measured today and used to test the evolution of the Universe.
The final analyses of the Planck satellite observations of the cosmic microwave background (CMB) \cite{Aghanim:2018eyx} also determine with great precision the parameters of the $\Lambda$CDM model, including the effective number of neutrino species, $N_{\rm eff}$, and the
contribution to the total energy budget of the Universe provided by baryons, $\omega_b \equiv \Omega_b h^2$.
The precise measurement of these quantities---in particular of $\omega_b$, which is constrained with a precision better than 1\% in \cite{Aghanim:2018eyx}---implies that standard BBN can be treated as an effectively parameter-free model.
To this end, it also helps the precise measurement of the neutron lifetime $\tau_n = (879.4\pm 0.6)\,\mathrm{s}$ \cite{Zyla:2020zbs}.

In order to obtain constraints, BBN analyses must employ
the abundances of primordial light nuclei, especially $^{2}$H and $^{4}$He.
These abundances are obtained from the observation of pristine astrophysical environments, and most of their experimental values are in good agreement with the theoretical expectations of both $\Lambda$CDM and the standard model of particle physics.
The only exception is that of $^{7}$Li, whose theoretical prediction is a factor \mbox{$\sim$ 2--3} higher than the observed value.
While it seems unlikely that the lithium problem can be solved in the framework of nuclear physics \cite{Iliadis:2020jtc,Broggini:2012rk}, it could be produced by stellar depletion \cite{Korn:2006tv}. Note, however, that the interstellar lithium yield measured in the Small Magellanic Cloud is nearly equal to the BBN prediction, and in agreement with its value in stars older than 4 billion years \cite{Howk:2012rb}.

Experimental and observational improvements give BBN excellent sensitivity to early-Universe effects from new physical scenarios. However, in order to transform such sensitivity into a useful tool to make predictions, it is necessary to achieve a high precision both on the theoretical calculations and on the experimental measurements of the primordial yields.
Over the years, several studies contributed to refine BBN analyses.
The improvements involved several different aspects both on
the theoretical side
(improved estimate of the neutron to proton weak conversion rates \cite{Dicus:1982bz,Cambier:1982pc,Lopez:1998vk,Esposito:1998rc,Brown:2000cp}, detailed neutrino decoupling \cite{Froustey:2020mcq,Akita:2020szl,Bennett:2020zkv}, careful analysis of key reactions in the BBN nuclear network  \cite{Pisanti:2020efz,Inesta:2017jhc,Iliadis:2016vkw})
as well as on the experimental one
(where future improvements are expected from the use of 30 m class telescopes for deuterium \cite{Cooke:2016rky} and from new observations of the damping tail of the CMB acoustic peaks for $^4$He measurements \cite{Ichikawa:2007js}).

From a theoretical perspective, the calculation of the nuclei abundances relies on a solid understanding of the nuclear processes involved in the light elements' production chain.
The main reactions that control the production of light nuclei are those affecting $^{2}$H and $^{3}$He, since they are the lightest elements that form from the free nucleon bath of the early Universe, hence participating at the earliest stages of BBN. Therefore, a precise experimental measurement of the rates of these nuclear reactions is needed to guarantee a precise estimate of their primordial abundances. In this vein, the reaction $^2$H(p,$\gamma)^3$He, which was one of the main sources of uncertainties in the determination of the deuterium abundance, has been recently measured \cite{Mossa:2020qgj,Mossa:2020gjc} by the LUNA collaboration \cite{Cavanna:2019pme} with a $\sim 3\%$ precision in the relevant center of mass energy for BBN.

Historically, a theoretical description of BBN had its origins in the studies of \cite{Wagoner:1966pv,Wagoner:1969,Wagoner:1972jh,Kawano:1988vh,Smith:1992yy}. After the pioneering work in \cite{Kawano:1988vh}, updated public codes are available today to the purpose of facilitating BBN calculations: \pth \cite{Pisanti:2007hk,Consiglio:2017pot}, \texttt{AlterBBN} \cite{Arbey:2011nf,Arbey:2018zfh}, \texttt{PRIMAT} \cite{Pitrou:2018cgg}.
The code whose update we present in this paper, \pth \cite{Pisanti:2007hk}, was significantly improved to a new version (\pthdue) in \cite{Consiglio:2017pot}.
The main changes included a more portable set of libraries based on the \texttt{ODEPACK} package\footnote{See the URL: \url{http://computation.llnl.gov/casc/odepack/}.}, which substituted the previously used NAG libraries\footnote{See the URL: \url{https://www.nag.com/content/nag-library}.}, and the delivery of a Graphic User Interface (GUI) to facilitate the use of the code.
In the update presented in this paper, \pthnew, we improve the code in two main aspects. On the one side we update the main reactions responsible for deuterium synthesis in light of the recent results described in \cite{Mossa:2020gjc} and \cite{Pisanti:2020efz}, as reported in Section \ref{s:rates}. On the other side, we release a new version of the GUI, with improved running performances and new plotting functionalities, as detailed in Section \ref{s:newGUI}. We conclude the paper in Section \ref{s:concl}.





\section{Update of nuclear rates}
\label{s:rates}

The accurate evaluation of the nuclear rates and uncertainties implemented in a nucleosynthesis code is the prerequisite for a precise determination of the primordial abundances. Such an evaluation has been reported in details in \cite{Pisanti:2020efz}, to which we address the interested reader. Here we report on the implementation of these new rates in \pth together with the other main changes with respect to its previous version \cite{Consiglio:2017pot}.

There are three reactions that have been re-analised by employing the method introduced in \cite{Serpico:2004gx}: they are the $^2$H(p,$\gamma)^3$He  radiative capture and the deuteron-deuteron transfer reactions, $^2$H(d, n)$^3$He and $^2$H(d, p)$^3$H ($dp\gamma$, $ddn$ and $ddp$ in short).
These reactions are those that mainly contribute to the total uncertainty budget on primordial deuterium.
The first one, in particular, had an uncertainty of about $6\div10\%$ in the energy range relevant for BBN \cite{DiValentino:2014cta}, before its measurement by the LUNA collaboration.
The new data included in the analysis (in addition to the ones already considered for the previous fit released with \pthdue) are presented in \cite{Tisma:2019acf,Mossa:2020gjc} for the $dp\gamma$ reaction, and \cite{Tumino:2014} for the $ddn$ and $ddp$ reactions. We address the reader to the analysis reported in \cite{Pisanti:2020efz} for the detailed discussion of the new results and their comparison with other works in the literature. We only summarize here the main points.

\begin{figure}[p]
\begin{center}
\includegraphics[width=.6\textwidth]{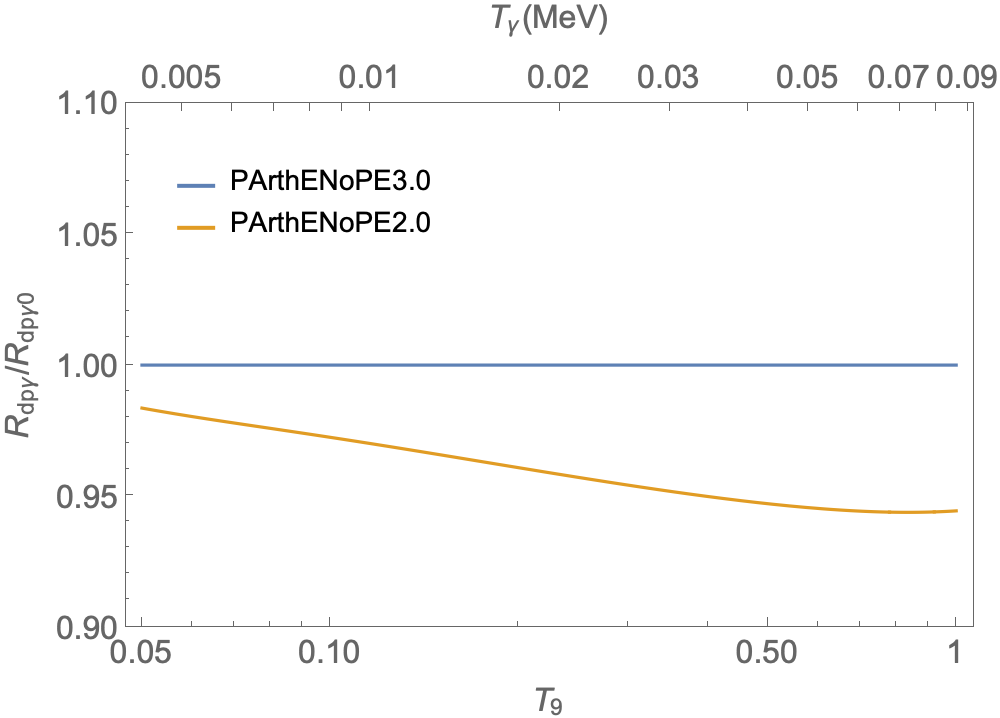}
\includegraphics[width=.6\textwidth]{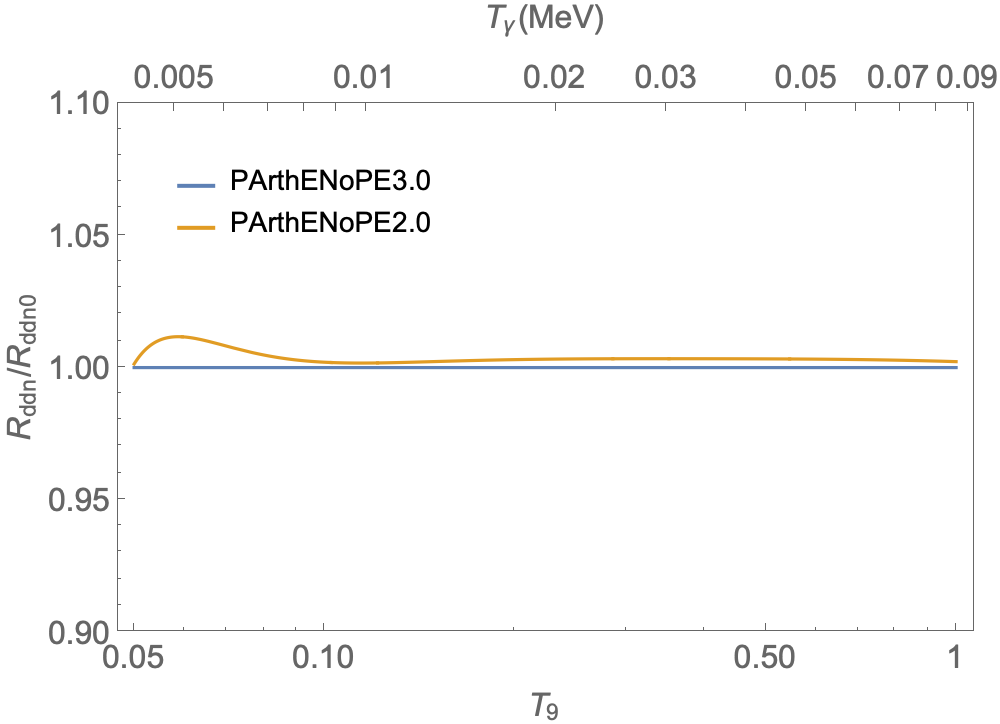}
\includegraphics[width=.6\textwidth]{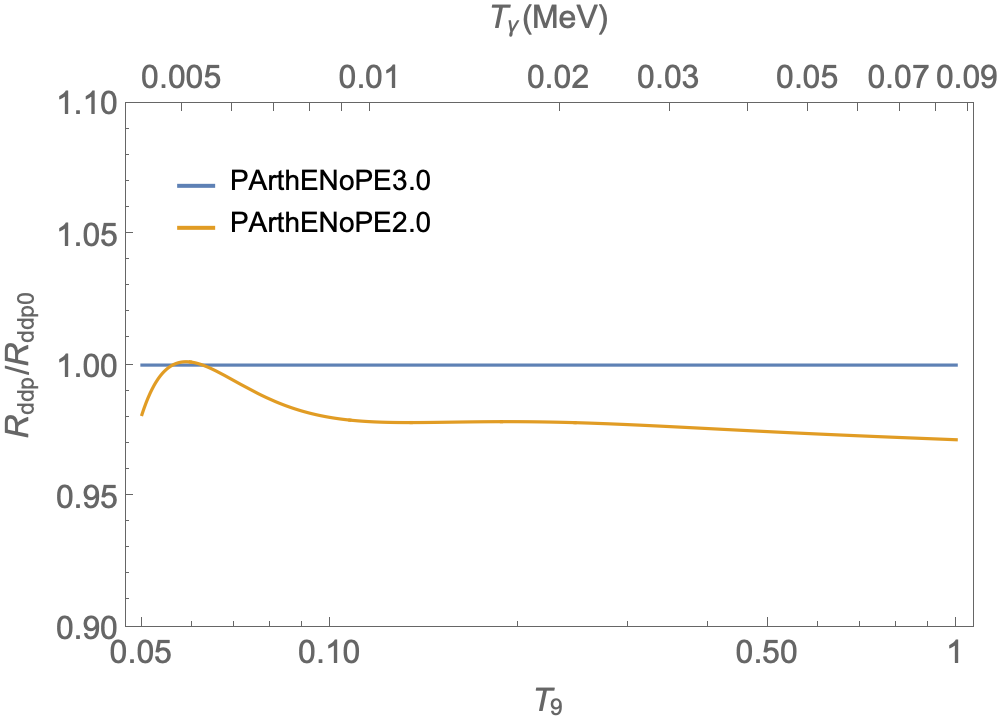}
\caption{From top to bottom, the ratios between the $dp\gamma$, $ddn$ and $ddp$ rates of \pthdue and \pthnew versus the temperature in $10^9$ K.}
\label{fig:ratios}
\end{center}
\end{figure}

We report in Fig.~\ref{fig:ratios} the comparison between the $dp\gamma$, $ddn$ and $ddp$ rates of \pthdue and \pthnew.
Both the $dp\gamma$ and $ddp$ rates show an increase with respect to their previous value, while $ddn$ is almost unchanged.
This translates in a decrease in the predicted abundance of deuterium, from $^2$H/H = $\left( 2.58 \pm 0.09\pm 0.03\right) \times 10^{-5}$ to $^2$H/H = $\left( 2.51 \pm 0.06\pm 0.03\right) \times 10^{-5}$, where we use $\omega_b = 0.02242\pm 0.00014$ \cite{Aghanim:2018eyx} and the two errors are obtained by propagation of the uncertainties on nuclear rates and baryon density.
The theoretical $^4$He mass fraction, $Y_p$, is instead almost the same, changing from $Y_p = 0.24684\pm 0.00004\pm 0.00012$ to $Y_p = 0.24687\pm 0.00003\pm 0.00012$.
The dominant contribution to the uncertainty, 0.00012, comes from the one on the neutron lifetime, $\tau_n$, for which we adopt the most recent average and error $\tau_n = (879.4 \pm 0.6)$~s \cite{Zyla:2020zbs}.

The new rates implemented in \pth, with their low and high 1-$\sigma$ variations as functions of the temperature $T_9\equiv T/(10^9\,\rm{K})$, are reported in \ref{app:A1}.
Note that the high and low rate values give the total uncertainty band on the thermal rates, as described in detail in \cite{Serpico:2004gx}.
They are obtained from the estimated error coming from the $\chi^2$ analysis, by inflating it by the usual factor $\sqrt{\chi^2}$ and summing in quadrature the overall scale error for the given reaction.

We have also improved the \pth algorithm itself in some aspects, for example by increasing the figures for the quantities printed during evolution, and correcting minor bugs, irrelevant from the point of view of the final numerical results, that only affected the printed output.

Most importantly, the user can now choose different determinations for the rates of the three deuterium destruction reactions mentioned above, specifying them in an auxiliary file, \texttt{rates.dat}.
The default values for these three reactions (PIS2020) are selected in case no such file is present in the working directory of \pthnew.
Table \ref{t:rates} shows the possible values for the three keywords DPG, DDN, and DDP, with the references that correspond to them. It is worth noting that the $dp\gamma$ rate PIS2020 is very similar to the LUNA2019 one, the only difference being the inclusion of the (almost irrelevant) data in \cite{Tisma:2019acf}, and that the DD rate PIS2020noTH do not include the Trojan Horse data of \cite{Tumino:2014}.

Notice that in order to run the Fortran code, either from command line or from within the GUI, the executable must be compiled. For ease of use, such a command has been included in the \texttt{Makefile} prepared for the GUI installation and can be invoked by using \texttt{make} without any option. 
The \texttt{Makefile} is configured to use the \texttt{gfortran} compiler as default, but the user can
modify the first line of the \texttt{Makefile} to indicate 
a different compiler.
A modification of the variable \texttt{FLIBS} in the \texttt{Makefile} is also required if the standard Fortran libraries are not available in the default paths, for example \texttt{FLIBS=-L/Library/Developer/CommandLineTools/SDKs/MacOSX.sdk/ usr/lib}.

\begin{table}[t]
\begin{center}
\begin{tabular}{cc}
\hline
DPG & DDN/DDP \\
\hline
PIS2020 \cite{Pisanti:2020efz} & PIS2020 \cite{Pisanti:2020efz} \\
LUNA2019 \cite{Mossa:2020gjc} & PIS2020noTH \cite{Pisanti:2020efz} \\
IL2016 \cite{Iliadis:2016vkw} & GI2017 \cite{Inesta:2017jhc} \\
MARCII \cite{Marcucci:2015yla} & COC2015 \cite{Coc:2015bhi} \\
COC2015 \cite{Coc:2015bhi} & PIS2007 \cite{Pisanti:2007hk} \\
MARCI \cite{Marcucci:2005zc} & CY2004 \cite{Cyburt:2004cq} \\
AD2011 \cite{Adelberger:2010qa} & \\
\hline
\end{tabular}
\end{center}
\caption{Values of the keywords, and corresponding reference, for the possible choices of the three rates $R_{dp\gamma}$, $R_{ddn}$, and $R_{ddp}$. }
\label{t:rates}
\end{table}

\section{The new GUI of \pth}
\label{s:newGUI}

The new \pth GUI has been almost completely rewritten from the previous version.
For the new GUI, we based all the graphical structure of the code on the Python package
\texttt{PySide2}, also known as \texttt{Qt for Python},
the official Python package for the \texttt{Qt 5} library%
\footnote{During the development of the new \pth GUI, a newer version of the package, named \texttt{PySide6} and based on \texttt{Qt 6}, was released. Nevertheless, we decided to let our code depend on \texttt{PySide2} in order to preserve support for Python 2.7.}.
This package allows to build a much more flexible and powerful GUI than the previously employed \texttt{Tkinter} package,
and it is completely portable for Python versions above 3.6.
Python 2.7 is still supported for Unix systems, while it does not work in Windows,
because of the \texttt{PySide2} requirements. Other Python packages that are required by the GUI are:
\texttt{appdirs} (default paths),
\texttt{matplotlib} (plotting functions) \cite{Hunter:2007ouj},
\texttt{numpy} (managing data and parameter grids) \cite{Harris:2020xlr},
\texttt{six} (Python 2 and Python 3 compatibility). 

To simplify the installation of the required packages, a simple \texttt{setup.py} file is provided.
The dependencies can be installed using
instructions of the \texttt{make} command, which must be executed from the same folder where the Makefile is located.
If the user does not employ virtual environments nor Anaconda, we recommend installing with \texttt{make installguiuser}, which stores the required dependencies in the local user paths.
Within a virtual environment, the previous command does not work and \texttt{make installgui} should be used instead.
Moreover, in some cases \texttt{sudo} \texttt{make} \texttt{installgui} can be required, if superuser privileges are needed.
These commands will use the default \texttt{python} version and the \texttt{PyPI} repositories to download and install the missing packages.
If the user wants to select a different python version, we recommend to edit the option \texttt{PY} at the beginning of the Makefile.
Alternatively, the user can specify it using for example \texttt{make installguiuser PY=python3} to use the specific \texttt{python3} instead of the default version.

For \texttt{Anaconda} users, we are aware of an incompatibility between the \texttt{PySide2} library available in \texttt{PyPI} and the \texttt{Qt} version installed by \texttt{Anaconda}.
Installing the dependencies using the \texttt{setup.py} or the \texttt{pip} command, therefore, is not a viable option: the command \texttt{make installconda} (or the bash script \texttt{conda.sh}) must be used instead.
Note that this command will create a new environment named ``parthenope3.0'', where all the dependencies will be installed using \texttt{conda install}.
If the user wants to change the environment name, there is the possibility to pass the new name through \texttt{make installconda CONDAENV=name} or \texttt{bash conda.sh name}, respectively.
Every time the user wants to open the GUI, in any case, the \texttt{Anaconda} environment must be activated.
This can be done with \texttt{conda activate parthenope3.0}, or using the chosen name.

In order to run the GUI, we provide the simple command \texttt{make rungui}, which is equivalent to use \texttt{python gui3.0.py \&} in its default version.
If the user wants to use a different python executable, it can be indicated in the Makefile as explained above, or the change can be explicited at runtime, for example using \texttt{make rungui PY=python3}.
Alternatively, one can directly run \texttt{python3 gui3.0.py \&}.

Since the GUI is composed of several fields that serve as input and the structure is quite complex, it may not work appropriately with small screen resolutions.
If the user cannot see the entire content of the window in the screen,
we recommend switching to a higher resolution.

Like the previous version of the GUI,
the new one is meant to simplify the creation of input cards and running the Fortran code.
The main improvements include the use of \texttt{Pool} from the \texttt{multiprocessing} package,
which allows to generate a queue of calls to the Fortran code that are launched exploiting all the available CPUs,
an improved control on the grid of input physical parameters for the Fortran code instances,
and the presence of new functions that allow the user to generate a number of useful plots (see below).

\subsection{Code structure}
The new GUI code is contained in one executable (\texttt{gui3.0.py})
and one python package (\texttt{parthenopegui}),
which contains a number of modules:
\begin{itemize}
\item \texttt{\_\_init\_\_.py}: in this file the version and content of the \texttt{parthenopegui} python package is defined.
\item \texttt{basic.py}: this file contains the initial configuration of the code and the functions used by the main executable.
\item \texttt{configuration.py}: it contains the functions and classes that are used
to ask/set, store and manage the parameters when they are prepared for the creation of the input cards.
\item \texttt{errorManager.py}: it contains the classes that show and save error messages into a log file (\texttt{gui.log}).
\item \texttt{mainWindow.py}: it contains the classes that define the main window of the GUI, creating the ``Home'' tab and importing the other ones.
\item \texttt{plotter.py}: it contains the classes that are used to read the output files of the Fortran code, store the relevant information in an organized way to produce plots, create the layout of the ``Plot'' tab and fill the figures.
\item \texttt{plotUtils.py}: it stores some utilities for the plots, some of which can be imported from exported scripts (see below).
\item \texttt{resourcesPySide2.py}, \texttt{resources.qrc}, \texttt{update\_resources.sh} and the \texttt{images/} folder: the icons and images used in the GUI buttons are stored in these files.
\item \texttt{runner.py}: it provides the functions and classes that connect the GUI with the Fortran code, by running the \pth executable and indicating the progress status in the GUI.
\item \texttt{runUtils.py}: it stores functions for preparing the input cards, compiling and running several instances of the code in parallel, and checking the status of the results. The functions in this file are imported also when running without the GUI (over ssh, for example).
\item \texttt{setrun.py}: it contains the classes that constitute the layout of the ``Run'' tab, which are used to configure all the parameters that enter the input card of the Fortran code and prepare the runs.
\item \texttt{tests.py}: it provides optional tests that check the consistency of all the functions used in the Python package and their correct behaviour. The test suite can be launched using the command \texttt{make testgui} to verify that everything is properly installed.
\item \texttt{texts.py}: it contains all the strings used in the code (error messages, warnings, instructions and tips).
\end{itemize}

\subsection{GUI structure}
Since the aim of the GUI is to simplify the use of the \pth Fortran code,
its layout is built in order to make immediately available to the user
most of the possible configurations and functions, in an intuitive way.
Most of the buttons and input fields have tooltips, which appear
when the mouse hovers for a few seconds on them, in order to further help the user to understand how to use the interface.
The main window is based on three tabs,
which are named ``Home'', ``Run'' and ``Plot''. The ``Home'' tab contains few information on the purpose of the code and a link to open the \pth web page. The other two tabs are described in the following subsections.

\subsubsection{The ``Run'' tab}
The ``Run'' tab contains all the controls for configuring the content of the input card and run the Fortran code. Its layout is presented in the \ref{app:A2}, where we provide some screen-shots of the GUI.

The top left panel allows us to select the size of the network of nuclides and reactions to be taken into account
(small, intermediate or complete, depending on the number of nuclides, which is 9, 18 or 26 respectively).

When a network size is selected, the reactions table (center left) is updated to contain the appropriate number of lines (40, 73 or 100 respectively).
Double clicking on the last column of each line allows the user to customize the reaction rate, changing it to the $\pm\sigma$ variations or adopting a different global normalization for rescaling the rate.

The panel on the bottom left controls the nuclides that constitute the output
of the Fortran code.
Only the evolution of the selected nuclides (left block) is saved in the output files,
while those appearing in the right block are allowed, given the current network size, but not selected.

In the bottom center, two more input fields are located:
one that controls the name of the folder
where the output files of the \pth Fortran code must be saved,
and one that activates or deactivates storing the full evolution of the abundances of each (selected) nuclide.
When the checkbox is not selected, only the final values will be available for further manipulation,
not the evolution.

The top right panel is where the user defines the series of points in the physical parameter space
that must be used in the execution of the Fortran code.
Single points as well as grids can be selected in a separate dialog window that is opened when clicking the ``Add a new point or grid'' button.
Within this dialog, one can easily select (column ``Type") between ``single values" or a ``grid" for each parameter. After closing this window, the selected values can be still modified using the ``Edit'' and ``Delete'' buttons that appear on the right of each grid/point summary table.
The total number of points, displayed just above the summary tables,
represents the number of independent calls to the Fortran code that will be executed when running.

The execution of the Fortran code is controlled by the bottom right buttons, which offer four options:
``Run with default parameters'', to execute the Fortran code only once with the default parameters and store the output in the selected folder;
``Run with custom parameters'', to execute the Fortran code once for each chosen point in the parameter space;
``Save default parameters'' and
``Save custom parameters'', to store the selected parameters in a file \texttt{settings.obj} inside the specified folder, for later use via command line.
These last two functions are meant to let the user configure the runs via the GUI,
and execute them in another pc (a cluster, for example), without the need of using the GUI again.
If \texttt{settings.obj} is stored in some \texttt{folder/}, the execution of the Fortran code without the GUI is later performed with a command like \texttt{python gui3.0.py folder/} (use the Python executable of your choice here).
Note that the same python version must be used to save and later read the \texttt{settings.obj} file, which is also created in the folder chosen for the output files after pressing one of the ``Run with'' buttons.

When \pth is called within the GUI by clicking one of the ``Run with default/custom parameters'' buttons, a progress bar shows the amount of performed runs with respect to the total.
The user also sees some information messages in the bottom right part of the tab.


\subsubsection{The ``Plot'' tab}
Most of the new functions of the GUI are included in the ``Plot'' tab, which contains the controls to read the output of the Fortran code and produce different kinds of figures. Its layout is presented in the \ref{app:A2}, together with the one of the ``Run'' tab.

The plotting functionalities need to read the existing output files from a previous \pth run. 
Therefore, before starting the creation of a plot, a valid folder path
must be passed through the button at the top left of the tab.
In details, the folder must contain a \texttt{settings.obj} file previously created by the GUI with the same Python version, and the output files generated by the Fortran code.

The GUI offers the possibility to produce three types of plots: \textit{Evolution}, \textit{1D dependence} and \textit{2D dependence}.
All of them can be configured via the radio buttons at the top left.
The Evolution button allows the user to plot how the element abundances (and other relevant quantities) change with respect to the temperature of the cosmic plasma.
The 1D and 2D dependence buttons allow plotting the final values of a given quantity as a function of one or two physical parameters.
Notice that, depending on the configuration of the grid, some of these plots may be unavailable.
For example, grids where only one physical parameter is varied cannot be used to produce ``2D dependence'' plots.

Depending on the plot type, the content of the left part of the tab changes, as detailed below.

\begin{figure}[t]
\begin{center}
\includegraphics[width=.5\textwidth]{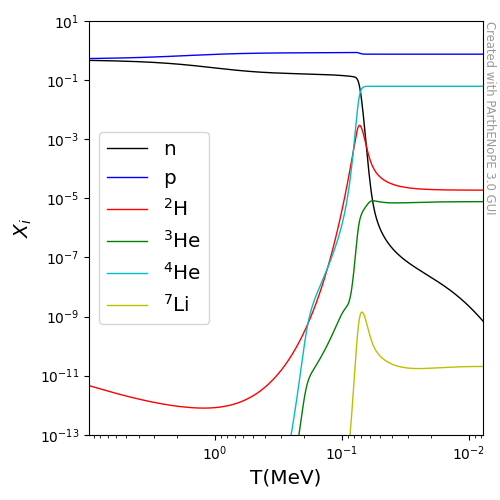}
\caption{Example plot produced with the ``Plot" tab of the GUI: evolution with temperature of the primordial abundances.}
\label{fig:evolution}
\end{center}
\end{figure}

When requesting ``Evolution'' plots, a table with all the available points in the grid appears just below the radio buttons for selecting the plot type.
By double clicking on a row, a new window appears. There, the user can select which parameter must be used to show its evolution in the figure, and define few properties for the line that will be plotted.
After completing the configuration and accepting the content of the dialog, a green tick notifies that the selected point is in use in the left upper table.
Moreover, a new row appears in the left bottom table, which summarizes the properties of each line currently configured in the plot.
For each row in this table, the properties of their corresponding lines are organised in columns.
By double-clicking on a column of this table, one can either modify the plot properties for the selected row, delete the corresponding line from the plot or reorder it in the table.
The lines are plotted in the figure in the same order that appears in the table.
This affects the overlapping of the lines and their order in the legend.
An example of ``Evolution'' plot is shown in figure \ref{fig:evolution}.

When selecting the ``1D dependence'' option, the structure of the interface remains the same as in the previous case,
except for the fact that instead of single points in the parameter space, the code groups several of them.
Within each group, represented by a row of the table, one physical parameter is varied while the others are kept fixed.
Selecting the group, one can produce a plot that shows the dependence of the selected quantity with respect to the varying parameter. The rest of the interface remains the same as described above.

\begin{figure}[t]
\begin{center}
\includegraphics[width=.8\textwidth]{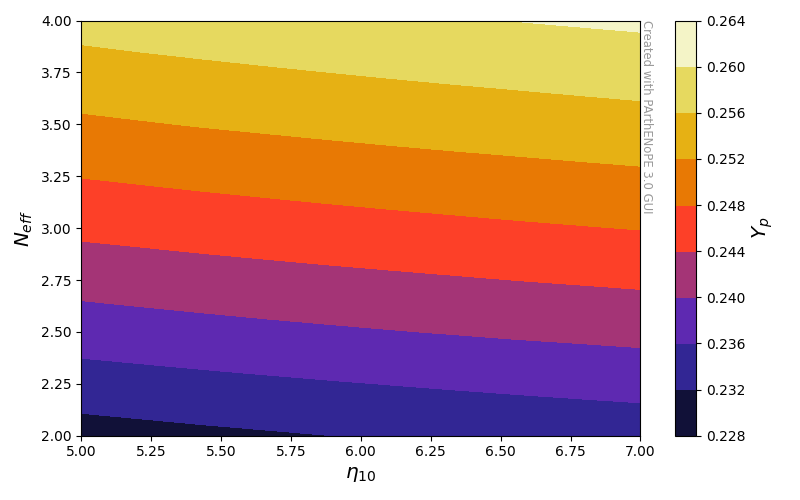}
\caption{Example plot produced by using the ``Plot" tab of the GUI: 2D color plot for $Y_p$ values as a function of $\eta_{10}$ and $N_\nu$.}
\label{fig:contourYp}
\end{center}
\end{figure}
For the ``2D dependence'' option, instead, the interface changes a bit.
Again, in the main table the points are grouped, this time two parameters are varying at the same time for each table row.
Since in this case the figure should contain contours instead of lines, the options that are available in the dialog window that appears after double-clicking on a table line are slightly different.
In the bottom table, in this case there is no option to reorder the contours, while the possibility to modify previously inserted settings or delete a contour are still available. See figure \ref{fig:contourYp} for an example.

\begin{figure}[p]
\begin{center}
\includegraphics[width=.5\textwidth]{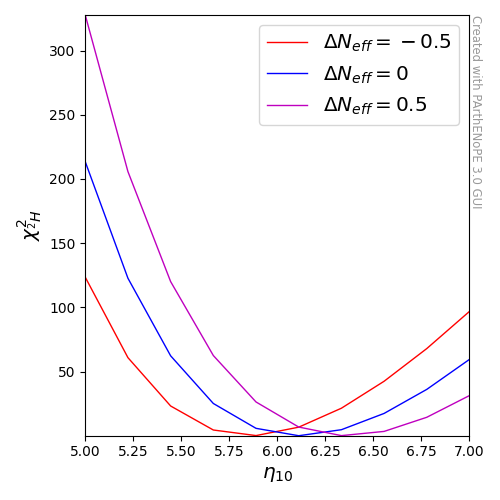}
\caption{Example plot produced by using the ``Plot" tab of the GUI: deuterium $\chi^2$ for three values of $\Delta N_\nu$ as a function of $\eta_{10}$.}
\label{fig:chi2D}
\end{center}
\end{figure}
\begin{figure}[p]
\begin{center}
\includegraphics[width=.7\textwidth]{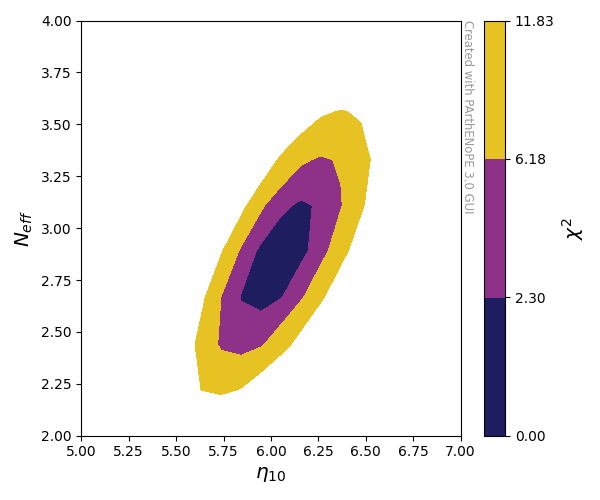}
\caption{Example plot produced by using the ``Plot" tab of the GUI: 2D contours in the $N_\nu$-$\eta_{10}$ plane at 1-, 2-, and 3-$\sigma$ C.L. obtained combining 
the $\chi^2$ of $^2$H/H and $Y_p$.}
\label{fig:contours}
\end{center}
\end{figure}
The interface also lets the user apply a Gaussian $\chi^2$ for producing both 1D and 2D plots.
In such case, the user has to input a mean and standard deviation in the window appearing after clicking on a data row, and the code will compute the Gaussian $\chi^2$ for the given quantity with respect to the varying parameter.
The user can also decide to plot the $\chi^2$, the likelihood or the log likelihood function.
A combination of the $\chi^2$ on several parameters can be also performed: for example, a typical BBN-only analysis considers the combination of deuterium and helium information.
Examples of the $\chi^2$ plots are shown in figure \ref{fig:chi2D} (1D example for one parameter) and figure \ref{fig:contours} (2D example for two parameters).

In the panel at the bottom right, the user has a preview of the final image, which is updated automatically when the content is changed. Several settings (axis labels, scales, limits, legend settings and more) can be adjusted using the three tabs above the figure: ``Axes settings", ``Legend settings", and ``Figure settings". In particular, in the axis and legend labels the user can use \LaTeX{} syntax to have the names properly rendered.
Notice that, because of space constraints in the visible window, some of the settings cannot be applied to the figure shown in the GUI and the figure shape and appearance do not necessarily reflect the true image content. However, the correct appearance is guaranteed in the exported image (see below).

Five buttons just above the figure allow the user to, respectively, refresh the image content (after having modified some figure properties); revert the changes made by the user on the plot settings to the default values; reset the image content (this will revert the user changes on the plot properties and delete the existing plot content); save a copy of the image in an external file; or export the python script to generate the plots outside of the GUI, in case one wants to further customize the obtained figures. As an example, we report in \ref{app:A3} the script corresponding to figure~\ref{fig:contourYp}.

\section{Conclusions}
\label{s:concl}

BBN has evolved from being one of the pillars of the standard Big Bang scenario to one of the most useful consistency tools to test the standard cosmological model.
\pth is a public tool developed since 2007 with the goal of helping researchers interested in this field.
The code fulfills the requirements of precision and versatility.
This updated version, in particular, represents an advance in both these requirements: we have implemented updated nuclear rates for the most important reactions of deuterium destruction, $^2$H(p,$\gamma)^3$He, $^2$H(d, n)$^3$He and $^2$H(d, p)$^3$H, and re-designed the previous version of the GUI, extending its functionalities for an easy implementation of grid running and plot preparation, which we hope will be of some help to BBN researchers.

\section*{Acknowledgments}
We are grateful to many users of \pth, who during these years, gave important feedbacks and suggestions to improve the code. 
OP work was supported by the Italian grant 2017W4HA7S “NAT-NET: Neutrino and Astroparticle Theory Network” (PRIN 2017) funded by the Italian Ministero dell’Istruzione, dell’Università e della Ricerca (MIUR), and Iniziativa Specifica TAsP of INFN. 
PFdS acknowledges support by the Vetenskapsr{\aa}det (Swedish Research Council) through contract No.~638-2013-8993 and the Oskar Klein Centre for Cosmoparticle Physics.
SG acknowledges financial support by the ``Juan de la Cierva-Incorporaci\'on'' program (IJC2018-036458-I) of the Spanish MICINN, by the Spanish grants FPA2017-85216-P (AEI/FEDER, UE), PROMETEO/2018/165 (Generalitat Valenciana) and the Red Consolider MultiDark FPA2017-90566-REDC, until September 2020, and from the European Union's Horizon 2020 research and innovation programme under the Marie Skłodowska-Curie grant agreement No 754496 (project FELLINI) starting from October 2020.
SG also thanks the Institute for Nuclear Theory at the University of Washington for its hospitality and the Department of Energy for partial support during the preparation of this article.

\newpage 

\appendix

\section{Fits of the rates}
\label{app:A1}
In this section we collect the analytical expressions implemented in the new version of the code, \pthnew, for $R_{dp\gamma}$, $R_{ddn}$, and $R_{ddp}$, as functions of $T_9$, together with their $1\sigma$ low and high variations. These functions correspond to the choice PIS2020 for the keywords in Table \ref{t:rates}.
The accuracy of the expressions is better than 0.06\% on the range $0.01\leq T_9\leq 4.0$.

\begin{itemize}
\item 
\mbox{$^2$H(p, $\gamma)^3$He}
\bea
&& \!\!\!\!\!\!\!\!\!\!\!\!\!\!\!\!\!\!\!\!\!\!\!\!R_{dp\gamma}(T_9) = T_9\,^{-2/3}\, \rm{exp} \left( -\frac{1.29043}{T_9^{1/3}} \right) \nn \\
&& \!\!\!\!\!\!\!\!\!\!\!\!\!\!\!\!\!\!\!\big( -6.194453 + 2.017433\cdot 10^2\, T_9^{1/3} - 2.817405\cdot 10^3\, T_9^{2/3} + 2.217193\cdot 10^4\, T_9 \nn \\
&& \!\!\!\!\!\!\!\!\!\!\!\!\!\!\!\!\!\!\!- 1.090730\cdot 10^5\, T_9^{4/3} + 3.556680\cdot 10^5\, T_9^{5/3} - 8.125130\cdot 10^5\, T_9^2 \nn \\
&& \!\!\!\!\!\!\!\!\!\!\!\!\!\!\!\!\!\!\!+ 1.379646\cdot 10^6\, T_9^{7/3} - 1.781593\cdot 10^6\, T_9^{8/3} + 1.772252\cdot 10^6\, T_9^3 \nn \\
&& \!\!\!\!\!\!\!\!\!\!\!\!\!\!\!\!\!\!\!- 1.364133\cdot 10^6\, T_9^{10/3} + 8.106062\cdot 10^5\, T_9^{11/3} - 3.683510\cdot 10^5\, T_9^4 \nn \\
&& \!\!\!\!\!\!\!\!\!\!\!\!\!\!\!\!\!\!\!+ 1.256020\cdot 10^5\, T_9^{13/3} - 3.109532\cdot 10^4\, T_9^{14/3} + 5.275483\cdot 10^3\, T_9^5 \nn \\
&& \!\!\!\!\!\!\!\!\!\!\!\!\!\!\!\!\!\!\!- 5.484033\cdot 10^2\, T_9^{16/3} + 2.633404\cdot 10^1\, T_9^{17/3} \big) \nn \\
&& \!\!\!\!\!\!\!\!\!\!\!\!\!\!\!\!\!\!\!\!\!\!\!\!R_{\rm dp\gamma, low}(T_9) = T_9\,^{-2/3}\, \rm{exp} \left( -\frac{1.29043}{T_9^{1/3}} \right) \nn \\
&& \!\!\!\!\!\!\!\!\!\!\!\!\!\!\!\!\!\!\!\big( -2.509725 + 9.699233\cdot 10^1\, T_9^{1/3} - 1.501097\cdot 10^3\, T_9^{2/3} + 1.244791\cdot 10^4\, T_9 \nn \\
&& \!\!\!\!\!\!\!\!\!\!\!\!\!\!\!\!\!\!\!- 6.170178\cdot 10^4\, T_9^{4/3} + 1.937398\cdot 10^5\, T_9^{5/3} - 4.084678\cdot 10^5\, T_9^2 \nn \\
&& \!\!\!\!\!\!\!\!\!\!\!\!\!\!\!\!\!\!\!+ 6.238214\cdot 10^5\, T_9^{7/3} - 7.051226\cdot 10^5\, T_9^{8/3} + 5.95309\cdot 10^5\, T_9^3 \nn \\
&& \!\!\!\!\!\!\!\!\!\!\!\!\!\!\!\!\!\!\!- 3.742405\cdot 10^5\, T_9^{10/3} + 1.726787\cdot 10^5\, T_9^{11/3} - 5.679486\cdot 10^4\, T_9^4 \nn \\
&& \!\!\!\!\!\!\!\!\!\!\!\!\!\!\!\!\!\!\!+ 1.260333\cdot 10^4\, T_9^{13/3} - 1.690983\cdot 10^3\, T_9^{14/3} + 1.036004\cdot 10^2\, T_9^5 \big) \nn \\
&& \!\!\!\!\!\!\!\!\!\!\!\!\!\!\!\!\!\!\!\!\!\!\!\!R_{\rm dp\gamma, high}(T_9) = T_9\,^{-2/3}\, \rm{exp} \left( -\frac{1.29043}{T_9^{1/3}} \right) \nn \\
&& \!\!\!\!\!\!\!\!\!\!\!\!\!\!\!\!\!\!\!\big( -2.670306 + 1.031982\cdot 10^2\, T_9^{1/3} - 1.597143\cdot 10^3\, T_9^{2/3} + 1.324437\cdot 10^4\, T_9 \nn \\
&& \!\!\!\!\!\!\!\!\!\!\!\!\!\!\!\!\!\!\!- 6.564967\cdot 10^4\, T_9^{4/3} + 2.06136\cdot 10^5\, T_9^{5/3} - 4.34603\cdot 10^5\, T_9^2 \nn \\
&& \!\!\!\!\!\!\!\!\!\!\!\!\!\!\!\!\!\!\!+ 6.637357\cdot 10^5\, T_9^{7/3} - 7.502388\cdot 10^5\, T_9^{8/3} + 6.33399\cdot 10^5\, T_9^3 \nn \\
&& \!\!\!\!\!\!\!\!\!\!\!\!\!\!\!\!\!\!\!- 3.981857\cdot 10^5\, T_9^{10/3} + 1.837273\cdot 10^5\, T_9^{11/3} - 6.04288\cdot 10^5\, T_9^4 \nn \\
&& \!\!\!\!\!\!\!\!\!\!\!\!\!\!\!\!\!\!\!+ 1.340973\cdot 10^4\, T_9^{13/3} - 1.799178\cdot 10^3\, T_9^{14/3} + 1.102291\cdot 10^2\, T_9^5 \big) \nn
\eea

\item
$^2$H(d, n)$^3$He
\bea
&& \!\!\!\!\!\!\!\!\!\!\!\!\!\!\!\!\!\!\!\!\!\!\!\!R_{ddn}(T_9) = T_9\,^{-2/3}\, \rm{exp} \left( -\frac{1}{T_9^{1/3}} \right) \nn \\
&& \!\!\!\!\!\!\!\!\!\!\!\!\!\!\!\!10^6\cdot \bigg( 6.04006 - 87.0655\, T_9^{1/3} + 532.179\, T_9^{2/3} - 1789.78\, T_9 + 3646.07\, T_9^{4/3} \nn \\
&& \!\!\!\!\!\!\!\!\!\!\!\!\!\!\!\!- 4843.13\, T_9^{5/3} + 4594.34\, T_9^2 - 3108.11\, T_9^{7/3} + 1485.45\, T_9^{8/3} - 493.762\, T_9^3 \nn \\
&& \!\!\!\!\!\!\!\!\!\!\!\!\!\!\!\!+ 109.061\, T_9^{10/3} - 14.1017\, T_9^{11/3} + 0.739461\, T_9^4 \bigg) \nn \\
&& \!\!\!\!\!\!\!\!\!\!\!\!\!\!\!\!\!\!\!\!\!\!\!\!R_{\rm ddn, low}(T_9) = T_9\,^{-2/3}\, \rm{exp} \left( -\frac{1}{T_9^{1/3}} \right) \nn \\
&& \!\!\!\!\!\!\!\!\!\!\!\!\!\!\!\!10^7\cdot \bigg( 3.72541 - 51.1407\, T_9^{1/3} + 308.658\, T_9^{2/3} - 1080.20\, T_9 + 2442.53\, T_9^{4/3} \nn \\
&& \!\!\!\!\!\!\!\!\!\!\!\!\!\!\!\!- 3787.02\, T_9^{5/3} + 4175.06\, T_9^2 - 3296.16\, T_9^{7/3} + 1850.77\, T_9^{8/3} - 721.825\, T_9^3 \nn \\
&& \!\!\!\!\!\!\!\!\!\!\!\!\!\!\!\!+ 185.745\, T_9^{10/3} - 28.2884\, T_9^{11/3} + 1.92205\, T_9^4 \bigg) \nn
\eea
\bea
&& \!\!\!\!\!\!\!\!\!\!\!\!\!\!\!\!\!\!\!\!\!\!\!\!R_{\rm ddn, high}(T_9) = T_9\,^{-2/3}\, \rm{exp} \left( -\frac{1}{T_9^{1/3}} \right) \nn \\
&& \!\!\!\!\!\!\!\!\!\!\!\!\!\!\!\!10^7\cdot \bigg( -2.51740 + 33.7276\, T_9^{1/3} - 202.222\, T_9^{2/3} + 722.247\, T_9 - 1713.32\, T_9^{4/3} \nn \\
&& \!\!\!\!\!\!\!\!\!\!\!\!\!\!\!\!+ 2818.40\, T_9^{5/3} - 3256.20\, T_9^2 + 2674.54\, T_9^{7/3} - 1553.68\, T_9^{8/3} + 623.072\, T_9^3 \nn \\
&& \!\!\!\!\!\!\!\!\!\!\!\!\!\!\!\!- 163.932\, T_9^{10/3} + 25.4681\, T_9^{11/3} - 1.77416\, T_9^4 \bigg) \nn
\eea

\item
$^2$H(d, p)$^3$H
\bea
&& \!\!\!\!\!\!\!\!\!\!\!\!\!\!\!\!\!\!\!\!\!\!\!\!R_{ddp}(T_9) = T_9\,^{-2/3}\, \rm{exp} \left( -\frac{1}{T_9^{1/3}} \right) \nn \\
&& \!\!\!\!\!\!\!\!\!\!\!\!\!\!\!\!10^6\cdot \bigg( 0.0524300 - 15.4274\, T_9^{1/3} + 158.367\, T_9^{2/3} - 652.819\, T_9 + 1356.03\, T_9^{4/3} \nn \\
&& \!\!\!\!\!\!\!\!\!\!\!\!\!\!\!\!- 1557.58\, T_9^{5/3} + 1160.00\, T_9^2 - 558.512\, T_9^{7/3} + 168.251\, T_9^{8/3} - 29.0298\, T_9^3 \nn \\
&& \!\!\!\!\!\!\!\!\!\!\!\!\!\!\!\!+ 2.19212\, T_9^{10/3} \bigg) \nn \\
&& \!\!\!\!\!\!\!\!\!\!\!\!\!\!\!\!\!\!\!\!\!\!\!\!R_{\rm ddp, low}(T_9) = T_9\,^{-2/3}\, \rm{exp} \left( -\frac{1}{T_9^{1/3}} \right) \nn \\
&& \!\!\!\!\!\!\!\!\!\!\!\!\!\!\!\!10^6\cdot \bigg( 0.265188 - 17.8087\, T_9^{1/3} + 169.592\, T_9^{2/3} - 682.042\, T_9 + 1402.55\, T_9^{4/3} \nn \\
&& \!\!\!\!\!\!\!\!\!\!\!\!\!\!\!\!- 1605.11\, T_9^{5/3} + 1191.63\, T_9^2 - 572.304\, T_9^{7/3} + 172.062\, T_9^{8/3} - 29.6345\, T_9^3 \nn \\
&& \!\!\!\!\!\!\!\!\!\!\!\!\!\!\!\!+ 2.23394\, T_9^{10/3} \bigg) \nn \\
&& \!\!\!\!\!\!\!\!\!\!\!\!\!\!\!\!\!\!\!\!\!\!\!\!R_{\rm ddp, high}(T_9) = T_9\,^{-2/3}\, \rm{exp} \left( -\frac{1}{T_9^{1/3}} \right) \nn \\
&& \!\!\!\!\!\!\!\!\!\!\!\!\!\!\!\!10^6\cdot \bigg( -0.160328 - 13.0461\, T_9^{1/3} + 147.141\, T_9^{2/3} - 623.597\, T_9 \nn \\
&& \!\!\!\!\!\!\!\!\!\!\!\!\!\!\!\!+ 1309.52\, T_9^{4/3} - 1510.05\, T_9^{5/3} + 1128.37\, T_9^2 - 544.719\, T_9^{7/3} + 164.439\, T_9^{8/3} \nn \\
&& \!\!\!\!\!\!\!\!\!\!\!\!\!\!\!\!- 28.4252\, T_9^3 + 2.15030\, T_9^{10/3} \bigg) \nn
\eea

\end{itemize}


\section{GUI screen-shots}
\label{app:A2}
\begin{figure}[h]
\begin{center}
\includegraphics[width=.7\textwidth]{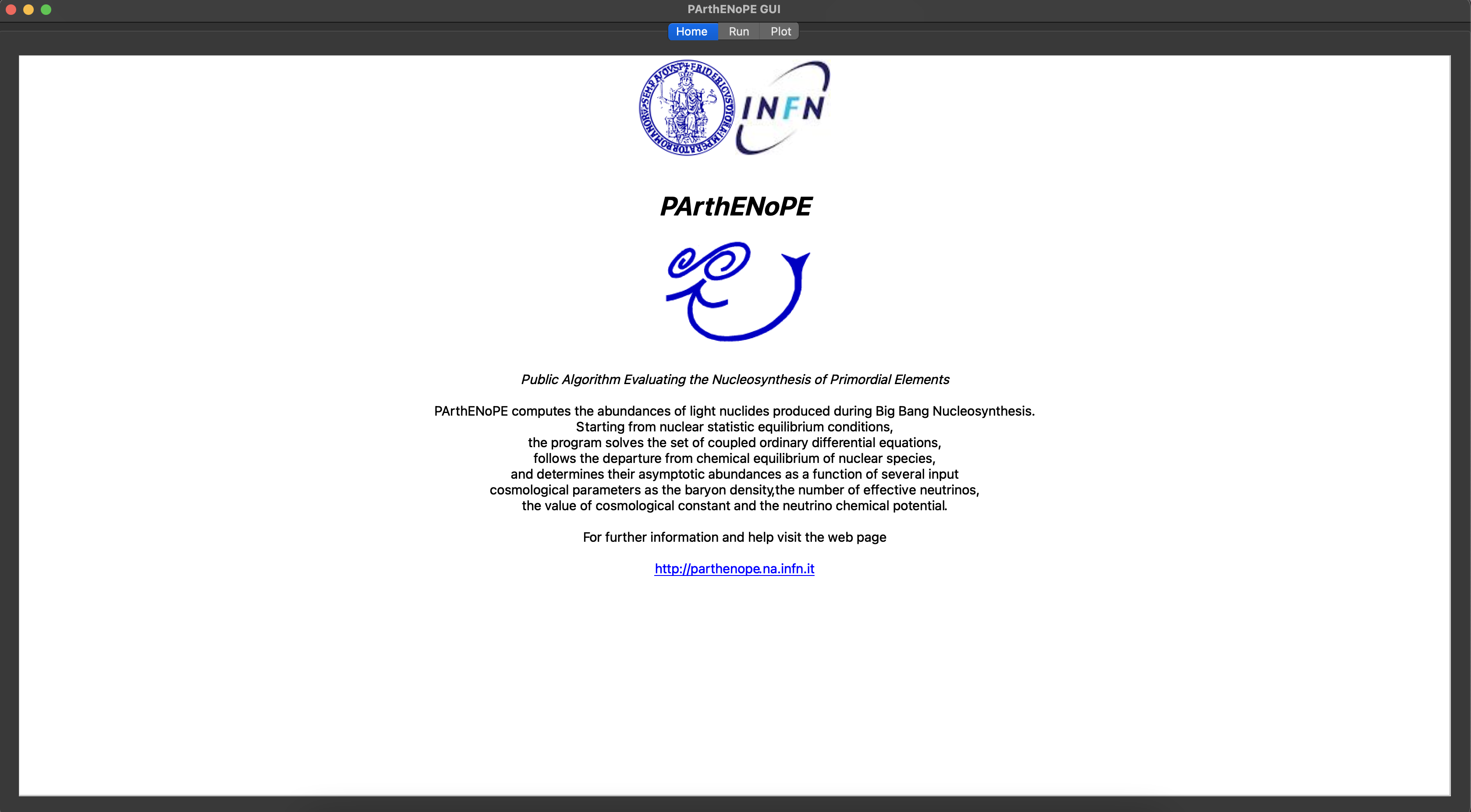}\\
\caption{``Home" tab of the GUI.}
\label{fig:GUI_windows_1}
\end{center}
\end{figure}

\begin{figure}[p]
\begin{center}
\includegraphics[width=.7\textwidth]{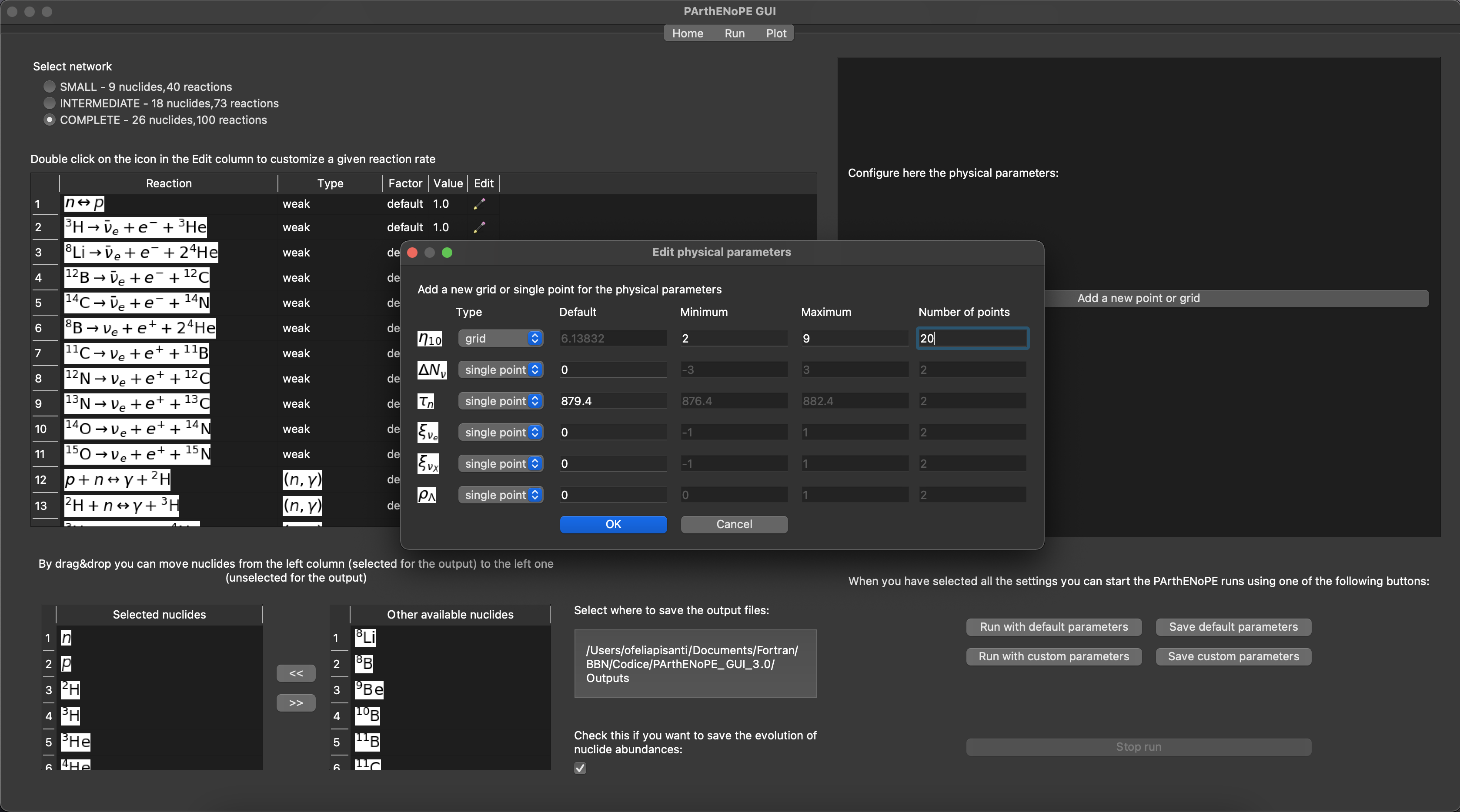}\\
\vspace{.3truecm}
\includegraphics[width=.7\textwidth]{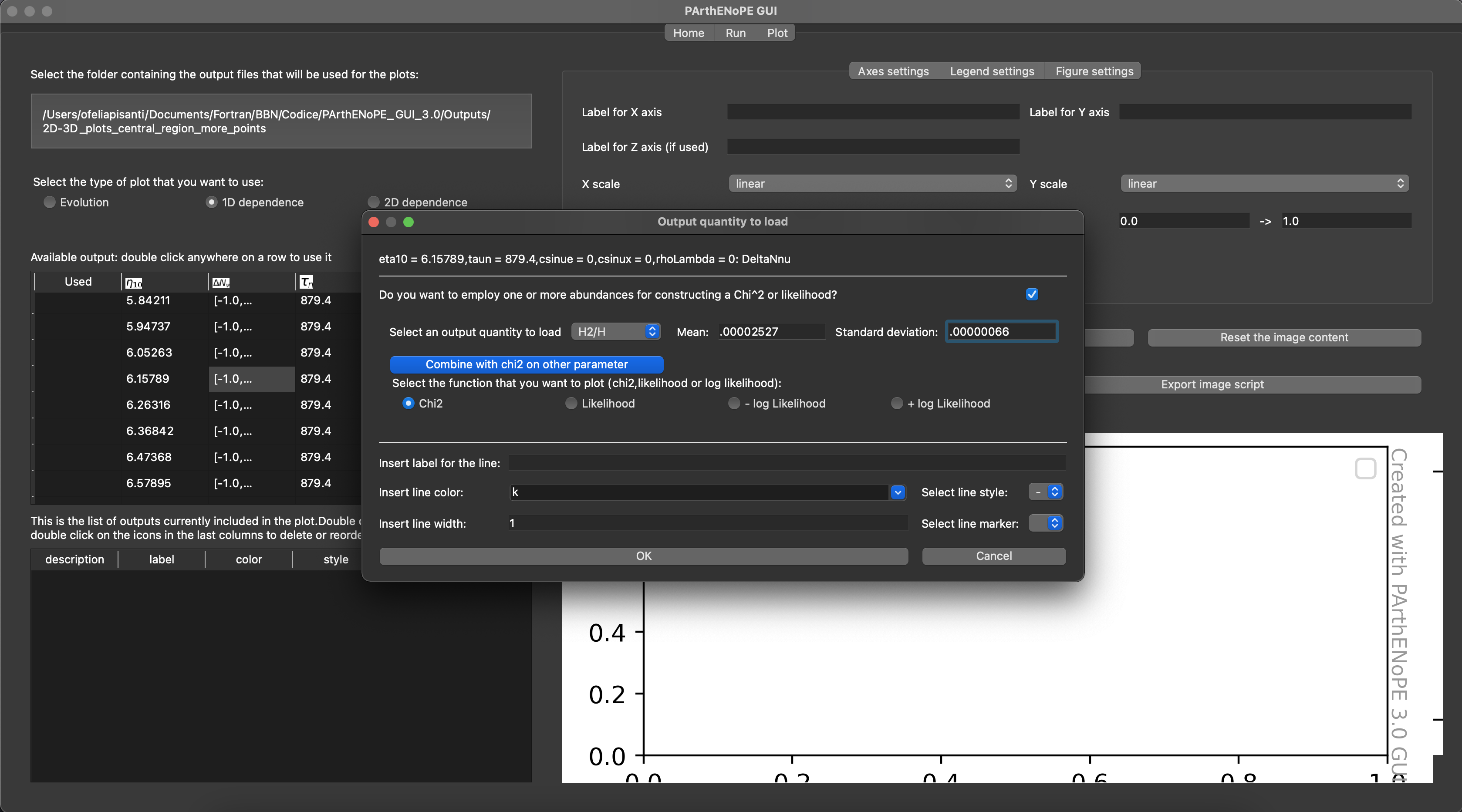}
\caption{\textit{Top panel}: ``Run'' tab of the GUI, with the dialog window that is opened when clicking on the button ``Add a new point or grid''. \textit{Bottom panel}: ``Plot'' tab of the GUI, with the dialog window that is opened when double-clicking on an available point in the center-left table to add it to the plot content.}
\label{fig:GUI_windows_2}
\end{center}
\end{figure}

We show here three screen-shots from the current version of the GUI on a MacOS system. Figure~\ref{fig:GUI_windows_1} shows the ``Home" tab of the GUI, while Figure~\ref{fig:GUI_windows_2} shows the layout of the ``Run'' (top panel) and ``Plot'' (bottom panel) tabs, together with a representative dialog window that is opened by performing an action. In the first case we can see the dialog window that permits the user to define a set of parameters for configuring the grid of runs,
while in the second one we show the window where a $\chi^2$ curve can be configured for the plot.

\newpage

\section{Example script produced by the GUI}
\label{app:A3}
We report here the python script corresponding to figure \ref{fig:contourYp}, produced with the specific button of the ``Plot" tab of the GUI.

\begin{verbatim}
"""File generated by the PArthENoPE GUI"""
import numpy as np
import matplotlib
matplotlib.use('agg')
import matplotlib.pyplot as plt
from parthenopegui.plotter import PGLine, PGContour

fig = plt.figure(figsize=(8.0, 5.0))
plt.plot(np.nan, np.nan)

lines = []

contours = []
contours.append(
    PGContour(
        np.asarray([
            5.0, 5.22222, 5.44444, 5.66667, 5.88889,
            6.11111, 6.33333, 6.55556, 6.77778, 7.0
        ]),
        np.asarray([2.0, 2.5, 3.0, 3.5, 4.0]),
        np.asarray([
            [0.230374, 0.230794, 0.23122 , 0.231599, 0.231985,
            0.232331, 0.232683, 0.233005, 0.233324, 0.233617],
            [0.2379 , 0.238325, 0.238736, 0.239133, 0.239499,
            0.239856, 0.240196, 0.240539, 0.240831, 0.24114 ],
            [0.244869, 0.245292, 0.245715, 0.246105, 0.246483,
            0.246835, 0.247176, 0.247513, 0.247817, 0.248123],
            [0.251352, 0.251782, 0.252209, 0.252588, 0.252967,
            0.253325, 0.253668, 0.253977, 0.254314, 0.254621],
            [0.257412, 0.257866, 0.258267, 0.258659, 0.259032,
            0.259389, 0.259728, 0.260058, 0.260372, 0.26068 ]
        ]),
        0,
        c2=False,
        cm='CMRmap',
        d=(
            'taun = 879.4, csinue = 0, csinux = 0, rhoLambda = 0:'
            + 'eta10, DeltaNnu - Y_p'
        ),
        ex='neither',
        f=True,
        hcb=True,
        l='',
        lvs=None,
        xl='eta10',
        yl='N_eff',
        zl='Y_p',
    )
)

for c in contours:
    if c.filled:
        func = plt.contourf
    else:
        func = plt.contour
    options = {}
    if c.levels is not None:
        options["levels"] = c.levels
    if c.extend is not None:
        options["extend"] = c.extend
    options["cmap"] = matplotlib.cm.get_cmap(c.cmap)
    CS = func(c.x, c.y, c.z, **options)
    if c.hascbar:
        cbar = plt.colorbar(CS)
        cbar.ax.set_ylabel('$Y_p$', fontsize='x-large')

plt.xlabel('$\eta_{10}$', fontsize='x-large')
plt.ylabel('$N_{eff}$', fontsize='x-large')
plt.xscale('linear')
plt.yscale('linear')
plt.xlim((5.0, 7.0))
plt.ylim((2.0, 4.0))
plt.tight_layout()
plt.text(
    1.0, 1.0,
    "Created with PArthENoPE 3.0 GUI",
    color="#999999", fontsize="small",
    ha="left", rotation=-90,
    transform=plt.gca().transAxes, va="top"
)
plt.savefig('fig_210212_150932.pdf')
plt.close()

\end{verbatim}

\bibliographystyle{elsarticle-num}
\bibliography{refs}

\begin{thebibliography}{10}
\expandafter\ifx\csname url\endcsname\relax
  \def\url#1{\texttt{#1}}\fi
\expandafter\ifx\csname urlprefix\endcsname\relax\def\urlprefix{URL }\fi
\expandafter\ifx\csname href\endcsname\relax
  \def\href#1#2{#2} \def\path#1{#1}\fi

\bibitem{Aghanim:2018eyx}
N.~Aghanim, et~al., {Planck 2018 results. VI. Cosmological parameters}, Astron.
  Astrophys. 641 (2020) A6.
\newblock \href {http://arxiv.org/abs/1807.06209} {\path{arXiv:1807.06209}},
  \href {https://doi.org/10.1051/0004-6361/201833910}
  {\path{doi:10.1051/0004-6361/201833910}}.

\bibitem{Zyla:2020zbs}
P.~A. Zyla, et~al., {Review of Particle Physics}, PTEP 2020~(8) (2020) 083C01.
\newblock \href {https://doi.org/10.1093/ptep/ptaa104}
  {\path{doi:10.1093/ptep/ptaa104}}.

\bibitem{Iliadis:2020jtc}
C.~Iliadis, A.~Coc, {Thermonuclear reaction rates and primordial
  nucleosynthesis}, Astrophys. J. 901~(2) (2020) 127.
\newblock \href {http://arxiv.org/abs/2008.12200} {\path{arXiv:2008.12200}},
  \href {https://doi.org/10.3847/1538-4357/abb1a3}
  {\path{doi:10.3847/1538-4357/abb1a3}}.

\bibitem{Broggini:2012rk}
C.~Broggini, L.~Canton, G.~Fiorentini, F.~L. Villante, {The cosmological 7Li
  problem from a nuclear physics perspective}, JCAP 06 (2012) 030.
\newblock \href {http://arxiv.org/abs/1202.5232} {\path{arXiv:1202.5232}},
  \href {https://doi.org/10.1088/1475-7516/2012/06/030}
  {\path{doi:10.1088/1475-7516/2012/06/030}}.

\bibitem{Korn:2006tv}
A.~J. Korn, F.~Grundahl, O.~Richard, P.~S. Barklem, L.~Mashonkina, R.~Collet,
  N.~Piskunov, B.~Gustafsson, {A probable stellar solution to the cosmological
  lithium discrepancy}, Nature 442 (2006) 657--659.
\newblock \href {http://arxiv.org/abs/astro-ph/0608201}
  {\path{arXiv:astro-ph/0608201}}, \href {https://doi.org/10.1038/nature05011}
  {\path{doi:10.1038/nature05011}}.

\bibitem{Howk:2012rb}
J.~C. Howk, N.~Lehner, B.~D. Fields, G.~J. Mathews, {The detection of
  interstellar lithium in a low-metallicity galaxy}, Nature 489 (2012) 121.
\newblock \href {http://arxiv.org/abs/1207.3081} {\path{arXiv:1207.3081}},
  \href {https://doi.org/10.1038/nature11407} {\path{doi:10.1038/nature11407}}.

\bibitem{Dicus:1982bz}
D.~A. Dicus, E.~W. Kolb, A.~M. Gleeson, E.~C.~G. Sudarshan, V.~L. Teplitz,
  M.~S. Turner, {Primordial Nucleosynthesis Including Radiative, Coulomb, and
  Finite Temperature Corrections to Weak Rates}, Phys. Rev. D 26 (1982) 2694.
\newblock \href {https://doi.org/10.1103/PhysRevD.26.2694}
  {\path{doi:10.1103/PhysRevD.26.2694}}.

\bibitem{Cambier:1982pc}
J.-L. Cambier, J.~R. Primack, M.~Sher, {Finite Temperature Radiative
  Corrections to Neutron Decay and Related Processes}, Nucl. Phys. B 209 (1982)
  372, [Erratum: Nucl.Phys.B 222, 517--517 (1983)].
\newblock \href {https://doi.org/10.1016/0550-3213(83)90550-3}
  {\path{doi:10.1016/0550-3213(83)90550-3}}.

\bibitem{Lopez:1998vk}
R.~E. Lopez, M.~S. Turner, {An Accurate Calculation of the Big Bang Prediction
  for the Abundance of Primordial Helium}, Phys. Rev. D 59 (1999) 103502.
\newblock \href {http://arxiv.org/abs/astro-ph/9807279}
  {\path{arXiv:astro-ph/9807279}}, \href
  {https://doi.org/10.1103/PhysRevD.59.103502}
  {\path{doi:10.1103/PhysRevD.59.103502}}.

\bibitem{Esposito:1998rc}
S.~Esposito, G.~Mangano, G.~Miele, O.~Pisanti, {Precision rates for nucleon
  weak interactions in primordial nucleosynthesis and He-4 abundance}, Nucl.
  Phys. B 540 (1999) 3--36.
\newblock \href {http://arxiv.org/abs/astro-ph/9808196}
  {\path{arXiv:astro-ph/9808196}}, \href
  {https://doi.org/10.1016/S0550-3213(98)00757-3}
  {\path{doi:10.1016/S0550-3213(98)00757-3}}.

\bibitem{Brown:2000cp}
L.~S. Brown, R.~F. Sawyer, {Finite temperature corrections to weak rates prior
  to nucleosynthesis}, Phys. Rev. D 63 (2001) 083503.
\newblock \href {http://arxiv.org/abs/astro-ph/0006370}
  {\path{arXiv:astro-ph/0006370}}, \href
  {https://doi.org/10.1103/PhysRevD.63.083503}
  {\path{doi:10.1103/PhysRevD.63.083503}}.

\bibitem{Froustey:2020mcq}
J.~Froustey, C.~Pitrou, M.~C. Volpe, {Neutrino decoupling including flavour
  oscillations and primordial nucleosynthesis}, JCAP 12 (2020) 015.
\newblock \href {http://arxiv.org/abs/2008.01074} {\path{arXiv:2008.01074}},
  \href {https://doi.org/10.1088/1475-7516/2020/12/015}
  {\path{doi:10.1088/1475-7516/2020/12/015}}.

\bibitem{Akita:2020szl}
K.~Akita, M.~Yamaguchi, {A precision calculation of relic neutrino decoupling},
  JCAP 08 (2020) 012.
\newblock \href {http://arxiv.org/abs/2005.07047} {\path{arXiv:2005.07047}},
  \href {https://doi.org/10.1088/1475-7516/2020/08/012}
  {\path{doi:10.1088/1475-7516/2020/08/012}}.

\bibitem{Bennett:2020zkv}
J.~J. Bennett, G.~Buldgen, P.~F. de~Salas, M.~Drewes, S.~Gariazzo, S.~Pastor,
  Y.~Y.~Y. Wong, {Towards a precision calculation of $N_{\rm eff}$ in the
  Standard Model II: Neutrino decoupling in the presence of flavour
  oscillations and finite-temperature QED} (12 2020).
\newblock \href {http://arxiv.org/abs/2012.02726} {\path{arXiv:2012.02726}}.

\bibitem{Pisanti:2020efz}
O.~Pisanti, G.~Mangano, G.~Miele, P.~Mazzella, {Primordial Deuterium after
  LUNA: concordances and error budget} (11 2020).
\newblock \href {http://arxiv.org/abs/2011.11537} {\path{arXiv:2011.11537}}.

\bibitem{Inesta:2017jhc}
A.~I\~nesta G\'omez, C.~Iliadis, A.~Coc, {Bayesian estimation of thermonuclear
  reaction rates for deuterium+deuterium reactions}, Astrophys. J. 849~(2)
  (2017) 134.
\newblock \href {http://arxiv.org/abs/1710.01647} {\path{arXiv:1710.01647}},
  \href {https://doi.org/10.3847/1538-4357/aa9025}
  {\path{doi:10.3847/1538-4357/aa9025}}.

\bibitem{Iliadis:2016vkw}
C.~Iliadis, K.~Anderson, A.~Coc, F.~Timmes, S.~Starrfield, {Bayesian Estimation
  of Thermonuclear Reaction Rates}, Astrophys. J. 831~(1) (2016) 107.
\newblock \href {http://arxiv.org/abs/1608.05853} {\path{arXiv:1608.05853}},
  \href {https://doi.org/10.3847/0004-637X/831/1/107}
  {\path{doi:10.3847/0004-637X/831/1/107}}.

\bibitem{Cooke:2016rky}
R.~J. Cooke, M.~Pettini, K.~M. Nollett, R.~Jorgenson, {The primordial deuterium
  abundance of the most metal-poor damped Ly$\alpha$ system}, Astrophys. J.
  830~(2) (2016) 148.
\newblock \href {http://arxiv.org/abs/1607.03900} {\path{arXiv:1607.03900}},
  \href {https://doi.org/10.3847/0004-637X/830/2/148}
  {\path{doi:10.3847/0004-637X/830/2/148}}.

\bibitem{Ichikawa:2007js}
K.~Ichikawa, T.~Sekiguchi, T.~Takahashi, {Primordial Helium Abundance from CMB:
  a constraint from recent observations and a forecast}, Phys. Rev. D 78 (2008)
  043509.
\newblock \href {http://arxiv.org/abs/0712.4327} {\path{arXiv:0712.4327}},
  \href {https://doi.org/10.1103/PhysRevD.78.043509}
  {\path{doi:10.1103/PhysRevD.78.043509}}.

\bibitem{Mossa:2020qgj}
V.~Mossa, et~al., {Setup commissioning for an improved measurement of the
  D(p,$\gamma $)$^3$He cross section at Big Bang Nucleosynthesis energies: LUNA
  collaboration}, Eur. Phys. J. A 56~(5) (2020) 144.
\newblock \href {http://arxiv.org/abs/2005.00002} {\path{arXiv:2005.00002}},
  \href {https://doi.org/10.1140/epja/s10050-020-00149-1}
  {\path{doi:10.1140/epja/s10050-020-00149-1}}.

\bibitem{Mossa:2020gjc}
V.~Mossa, et~al., {The baryon density of the Universe from an improved rate of
  deuterium burning}, Nature 587~(7833) (2020) 210--213.
\newblock \href {https://doi.org/10.1038/s41586-020-2878-4}
  {\path{doi:10.1038/s41586-020-2878-4}}.

\bibitem{Cavanna:2019pme}
F.~Cavanna, {Nuclear Astrophysics at Gran Sasso Laboratory: the LUNA
  experiment}, CERN Proc. 1 (2019) 265--272.

\bibitem{Wagoner:1966pv}
R.~V. Wagoner, W.~A. Fowler, F.~Hoyle, {On the Synthesis of elements at very
  high temperatures}, Astrophys. J. 148 (1967) 3--49.
\newblock \href {https://doi.org/10.1086/149126} {\path{doi:10.1086/149126}}.

\bibitem{Wagoner:1969}
R.~V. {Wagoner}, {Synthesis of the Elements Within Objects Exploding from Very
  High Temperatures}, Astrophys. J. Suppl. 18 (1969) 247.
\newblock \href {https://doi.org/10.1086/190191} {\path{doi:10.1086/190191}}.

\bibitem{Wagoner:1972jh}
R.~V. Wagoner, {Big bang nucleosynthesis revisited}, Astrophys. J. 179 (1973)
  343--360.
\newblock \href {https://doi.org/10.1086/151873} {\path{doi:10.1086/151873}}.

\bibitem{Kawano:1988vh}
L.~Kawano, {Let's Go: Early Universe. Guide to Primordial Nucleosynthesis
  Programming} (3 1988).

\bibitem{Smith:1992yy}
M.~S. Smith, L.~H. Kawano, R.~A. Malaney, {Experimental, computational, and
  observational analysis of primordial nucleosynthesis}, Astrophys. J. Suppl.
  85 (1993) 219--247.
\newblock \href {https://doi.org/10.1086/191763} {\path{doi:10.1086/191763}}.

\bibitem{Pisanti:2007hk}
O.~Pisanti, A.~Cirillo, S.~Esposito, F.~Iocco, G.~Mangano, G.~Miele, P.~D.
  Serpico, {PArthENoPE: Public Algorithm Evaluating the Nucleosynthesis of
  Primordial Elements}, Comput. Phys. Commun. 178 (2008) 956--971.
\newblock \href {http://arxiv.org/abs/0705.0290} {\path{arXiv:0705.0290}},
  \href {https://doi.org/10.1016/j.cpc.2008.02.015}
  {\path{doi:10.1016/j.cpc.2008.02.015}}.

\bibitem{Consiglio:2017pot}
R.~Consiglio, P.~F. de~Salas, G.~Mangano, G.~Miele, S.~Pastor, O.~Pisanti,
  {PArthENoPE reloaded}, Comput. Phys. Commun. 233 (2018) 237--242.
\newblock \href {http://arxiv.org/abs/1712.04378} {\path{arXiv:1712.04378}},
  \href {https://doi.org/10.1016/j.cpc.2018.06.022}
  {\path{doi:10.1016/j.cpc.2018.06.022}}.

\bibitem{Arbey:2011nf}
A.~Arbey, {AlterBBN: A program for calculating the BBN abundances of the
  elements in alternative cosmologies}, Comput. Phys. Commun. 183 (2012)
  1822--1831.
\newblock \href {http://arxiv.org/abs/1106.1363} {\path{arXiv:1106.1363}},
  \href {https://doi.org/10.1016/j.cpc.2012.03.018}
  {\path{doi:10.1016/j.cpc.2012.03.018}}.

\bibitem{Arbey:2018zfh}
A.~Arbey, J.~Auffinger, K.~P. Hickerson, E.~S. Jenssen, {AlterBBN v2: A public
  code for calculating Big-Bang nucleosynthesis constraints in alternative
  cosmologies}, Comput. Phys. Commun. 248 (2020) 106982.
\newblock \href {http://arxiv.org/abs/1806.11095} {\path{arXiv:1806.11095}},
  \href {https://doi.org/10.1016/j.cpc.2019.106982}
  {\path{doi:10.1016/j.cpc.2019.106982}}.

\bibitem{Pitrou:2018cgg}
C.~Pitrou, A.~Coc, J.-P. Uzan, E.~Vangioni, {Precision big bang nucleosynthesis
  with improved Helium-4 predictions}, Phys. Rept. 754 (2018) 1--66.
\newblock \href {http://arxiv.org/abs/1801.08023} {\path{arXiv:1801.08023}},
  \href {https://doi.org/10.1016/j.physrep.2018.04.005}
  {\path{doi:10.1016/j.physrep.2018.04.005}}.

\bibitem{Serpico:2004gx}
P.~D. Serpico, S.~Esposito, F.~Iocco, G.~Mangano, G.~Miele, O.~Pisanti,
  {Nuclear reaction network for primordial nucleosynthesis: A Detailed analysis
  of rates, uncertainties and light nuclei yields}, JCAP 12 (2004) 010.
\newblock \href {http://arxiv.org/abs/astro-ph/0408076}
  {\path{arXiv:astro-ph/0408076}}, \href
  {https://doi.org/10.1088/1475-7516/2004/12/010}
  {\path{doi:10.1088/1475-7516/2004/12/010}}.

\bibitem{DiValentino:2014cta}
E.~Di~Valentino, C.~Gustavino, J.~Lesgourgues, G.~Mangano, A.~Melchiorri,
  G.~Miele, O.~Pisanti, {Probing nuclear rates with Planck and BICEP2}, Phys.
  Rev. D 90~(2) (2014) 023543.
\newblock \href {http://arxiv.org/abs/1404.7848} {\path{arXiv:1404.7848}},
  \href {https://doi.org/10.1103/PhysRevD.90.023543}
  {\path{doi:10.1103/PhysRevD.90.023543}}.

\bibitem{Tisma:2019acf}
I.~Ti\v{s}ma, M.~Lipoglav\v{s}ek, M.~Mihovilovi\v{c}, S.~Markelj, M.~Vencelj,
  J.~Vesi\'c, {Experimental cross section and angular distribution of
  the$^{2}$H(p, $\gamma$ )$^{3}$He reaction at Big-Bang nucleosynthesis
  energies}, Eur. Phys. J. A 55~(8) (2019) 137.
\newblock \href {https://doi.org/10.1140/epja/i2019-12816-1}
  {\path{doi:10.1140/epja/i2019-12816-1}}.

\bibitem{Tumino:2014}
A.~{Tumino}, R.~{Spart{\`a}}, C.~{Spitaleri}, A.~M. {Mukhamedzhanov},
  S.~{Typel}, R.~G. {Pizzone}, E.~{Tognelli}, S.~{Degl'Innocenti}, V.~{Burjan},
  V.~{Kroha}, Z.~{Hons}, M.~{La Cognata}, L.~{Lamia}, J.~{Mrazek}, S.~{Piskor},
  P.~G. {Prada Moroni}, G.~G. {Rapisarda}, S.~{Romano}, M.~L. {Sergi}, {New
  Determination of the $^{2}$H(d,p)$^{3}$H and $^{2}$H(d,n)$^{3}$He Reaction
  Rates at Astrophysical Energies}, Astrophys. J. 785~(2) (2014) 96.
\newblock \href {https://doi.org/10.1088/0004-637X/785/2/96}
  {\path{doi:10.1088/0004-637X/785/2/96}}.

\bibitem{Marcucci:2015yla}
L.~E. Marcucci, G.~Mangano, A.~Kievsky, M.~Viviani, {Implication of the
  proton-deuteron radiative capture for Big Bang Nucleosynthesis}, Phys. Rev.
  Lett. 116~(10) (2016) 102501, [Erratum: Phys.Rev.Lett. 117, 049901 (2016)].
\newblock \href {http://arxiv.org/abs/1510.07877} {\path{arXiv:1510.07877}},
  \href {https://doi.org/10.1103/PhysRevLett.116.102501}
  {\path{doi:10.1103/PhysRevLett.116.102501}}.

\bibitem{Coc:2015bhi}
A.~Coc, P.~Petitjean, J.-P. Uzan, E.~Vangioni, P.~Descouvemont, C.~Iliadis,
  R.~Longland, {New reaction rates for improved primordial D/H calculation and
  the cosmic evolution of deuterium}, Phys. Rev. D 92~(12) (2015) 123526.
\newblock \href {http://arxiv.org/abs/1511.03843} {\path{arXiv:1511.03843}},
  \href {https://doi.org/10.1103/PhysRevD.92.123526}
  {\path{doi:10.1103/PhysRevD.92.123526}}.

\bibitem{Marcucci:2005zc}
L.~E. Marcucci, M.~Viviani, R.~Schiavilla, A.~Kievsky, S.~Rosati,
  {Electromagnetic structure of A=2 and 3 nuclei and the nuclear current
  operator}, Phys. Rev. C 72 (2005) 014001.
\newblock \href {http://arxiv.org/abs/nucl-th/0502048}
  {\path{arXiv:nucl-th/0502048}}, \href
  {https://doi.org/10.1103/PhysRevC.72.014001}
  {\path{doi:10.1103/PhysRevC.72.014001}}.

\bibitem{Cyburt:2004cq}
R.~H. Cyburt, {Primordial nucleosynthesis for the new cosmology: Determining
  uncertainties and examining concordance}, Phys. Rev. D 70 (2004) 023505.
\newblock \href {http://arxiv.org/abs/astro-ph/0401091}
  {\path{arXiv:astro-ph/0401091}}, \href
  {https://doi.org/10.1103/PhysRevD.70.023505}
  {\path{doi:10.1103/PhysRevD.70.023505}}.

\bibitem{Adelberger:2010qa}
E.~G. Adelberger, et~al., {Solar fusion cross sections II: the pp chain and CNO
  cycles}, Rev. Mod. Phys. 83 (2011) 195.
\newblock \href {http://arxiv.org/abs/1004.2318} {\path{arXiv:1004.2318}},
  \href {https://doi.org/10.1103/RevModPhys.83.195}
  {\path{doi:10.1103/RevModPhys.83.195}}.

\bibitem{Hunter:2007ouj}
J.~D. Hunter, {Matplotlib: A 2D Graphics Environment}, Comput. Sci. Eng. 9~(3)
  (2007) 90--95.
\newblock \href {https://doi.org/10.1109/MCSE.2007.55}
  {\path{doi:10.1109/MCSE.2007.55}}.

\bibitem{Harris:2020xlr}
C.~R. Harris, et~al., {Array programming with NumPy}, Nature 585~(7825) (2020)
  357--362.
\newblock \href {http://arxiv.org/abs/2006.10256} {\path{arXiv:2006.10256}},
  \href {https://doi.org/10.1038/s41586-020-2649-2}
  {\path{doi:10.1038/s41586-020-2649-2}}.

\end{thebibliography}

\end{document}